# Evaluation of Non-Fungible Token (NFT)

**Priyanshu Lohar[1], Kiran Rathi[2]**

[1,2] Department of Electronics & Communiction, Swami Keshvanand Inst.of Tech, M & G, Jaipur 302017, India

Email: priyanshulohar1@gmail.com, kiran.rathi@skit.ac.in

**Abstract-** The derivative of token standard of Ethereum blockchain, termed as Non Fungible Token is distinguishable token. These tokens are bound with digital properties that provide them unique identification which helps in fulfilling the aim of distinguishable tokens. These tokens are used as an evidence of ownership for the digital asset, with which they are bound to. And it is with these non fungible tokens that the problem of proving ownership of digital asset is being solved and with this technique, it is with hope that developers are looking forward to solve many more problems of the real world with it, may it be providing tradability solutions for arts, real estate and many other sectors. During the time of writing this, the NFT has shown unpredictable growth in the recent years and this has caused the stimulation of prosperity of DApps(Decentralized Application).With an unpredictable growth and garnering attention worldwide with many mainstream key people investing in it , the NFT is still in developing stage and is still premature. This paper is an attempt to squeeze the NFT developments systematically, so the aspiring developers can have the resource to start with and aid the development process further.

## 1. Introduction

Non Fungible Token (NFT) is an Ethereum based token which uses smart contract functionality of ethereum to give unique identity to any digital asset, such as image, video,3d model, game punks, etc.[1]  NFT is a way to give unique identity to digital assets that will be the representation of ownership of that asset.[2] It  us tokenize the assets like art, collectible, etc ,as they are unique in nature and can only have one official owner at a time. The NFT is based on ethereum blockchain and hence no one can modify the record of ownership or make a duplicate copy of an existing NFT.[3]

Non Fungible is an economic term that is used to describe things that are not interchangeable for other items because of their unique properties. Fungible item on the other hand can be exchanged because their values define them and not the unique properties. For example, 1 rupee is exchangeable for another 1 rupee, but this is not the case with NFT as each one has an individual properties that are not like any other NFT that has been produced and this makes them unique. NFTs are paving way for the newer means of validation of distinctiveness in the modern cyber world.

Mainstream case of NFT purchase is Jack Dorsey, founder of square and twitter trading his first tweet for 2.9 million dollars and other such example is the artist Beeple selling his piece of art" Everyday: the First 5000 Days, for 69 million dollars. [4] These examples show the mainstream attention about the developing technology which might dominate the world in the coming future.

## 2. Technology Used

### 2.1 Blockchain

As the term suggests , it is comprised of concatenated blocks linked in such a way that no block can change its place, as changing it will affect the whole system and will cause a possible error which would be traceable and thus the technology provides a trust protocol in the digital world, which was the need of time, as the internet began to grow in the dimensions where banking, and storing legal data became a phenomenon that cannot be restored to the old way of recording and organizing the data in paper and arranging them with the help of some algorithm that only a librarian knows.

 As referenced over, the blockchain is a framework for amassing data in such pattern that is abstruse to modify or alter. It is a complex log of exchange that is scattered across the whole blockchain network. Each block in series has store of exchanges, and whenever new exchange happens its log is added to the network. It is a decentralized log of information visible to fellow user in network known as Distributed Ledger Technology (DLT), and it store exchange with cryptographic mark called hash [5].

 Blockchain is a peer to peer network of computers because of its decentralized architecture where users are directly interacting with each other without the need of trusted central intermediaries.

#### 2.1.1 Characteristics of Blockchain

Following are some properties anent NFT,
- *Decentralization*

As discussed previously, blockchain is a peer to peer network of computers because of its decentralized architecture it permits sharing digital assets all across the network. It quashes the potential single point of failure and doff of the need of central authority. [6]

- *Cryptographic Hashing*

Hashing is done in whole blockchain, which results in the formation of chronological chain. It ensures user of unaltered blockchain because altering of a specific hash value will lead to the altering of successive hash values and leading to an invalid chain formation. [7]

- *Time stamping*

Every record in blockchain sequentially gets time stamped. It renders trust characteristics like traceability, transparency, and full transaction history. Time stamping and cryptographic hash can be collated and used, for example, as a Proof of-Existence for certain data at a specific time.[5]

- *Immutability*

Cryptographic hashing and decentralized validation ensures that the blockchain is unaltered at any instant. Exceptions to this characteristic feature are attacks like the 51% - attack.[7]

- *Consensus Mechanism*

It defines how exchange or transaction are validated in blockchain network.[8] Since the development of Bitcoin blockchain and Proof-of-work, many unique operandi and coalition of existing consensus techniques are implemented in new blockchains.

- *Proof of Work*

Proof-of-Work is the way by which the nodes in the ledger network proves that they are a part of the network, for this they have to do heavy computations which verifies them to be the legal member of the network.[9] Proof-Of-Work provides immunity from the double spending problem & Sybil attack in decentralized network.

## 2.2 Ethereum

Ethereum is a decentralized, open source blockchain which has a special feature known as smart contract. It was introduced By Vitalik Buterin in a white paper in year 2013 where he addressed some limitation of the Bitcoin. The feature introduced in ethereum was turing-complete programming language which enables ethereum to support every calculation which includes loops also. Ethereum provides and abstract layer which enables the user to make their own rules for ownership, formats of transactions and state transition functions and also smart contracts, which are defined set of cryptographic rules that gets executed when certain conditions are fulfilled. [10]

For order sensitive execution most NFTs have reliance on smart contract.[4]

## 2.3 Transaction & Address

Blockchain address is an unique identifier for sending and receiving the digital assets, it comprises fixed number of alphanumeric character which are generated from the public key and the private key. If an owner wants to send an asset to another address then he has to prove the possession of the corresponding private key and then asset can be sent to another address with the correct digital signature. Cryptocurrency wallet is used to perform this operation using and this is based on ERC-777 standard.[3]

## 2.4 Data Encoding

In this operation the data is converted from one form into other. In Major Blockchains, like ethereum, the hex values are employed to encode the transaction credentials like, function names, parameters, return values. If someone claims that they own an NFT-based asset, he on literal ground owns the particular hex code which is associated with that asset. [3]

## 3. Basic Model of NFT System

This section will lay emphasis on the protocols, key properties and token standards of basic model of NFT system.

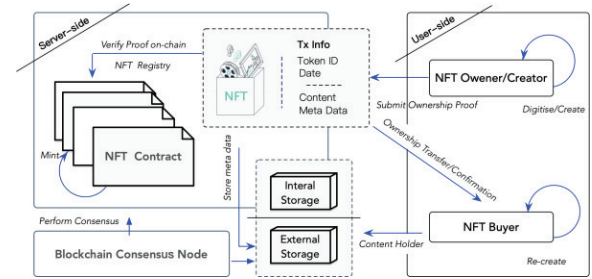

**Figure 1. Model of NFT System [3]**

The NFT framework is a blockchain based application with properties of ethereum used that are needed for the trade of digital assets digitally.

### 3.1 Protocol

NFT development requires distributed ledger for records & exchangeable transactions for trading in the peer to peer network. Here the distributed ledger is treated as special type of database, whose job is to store the NFT data. It is presumed for ledger to have basic security consistency, completeness and availability characteristic. As a platform for trading the NFT system has two onuses, the NFT buyer and the NFT owner.

The brief detail of the protocols of NFT system is as follows,

- *NFT Store*

External database is used by NFT owner to store the crude data. Storing data on internal blockchain is also allowed if user wants to [3].

171

- *NFT Sign.*

Owner of NFT integrates the digital signature with transactions which includes the hash of NFT data and then sending that transaction to smart contract [3].

- *NFT Digitize*

After checking that file, title, description and assuring that they are completely accurate, the user then digitizes the raw data into proper structured format [3].

- *NFT Mint and Trade*

When the smart contract receives the transactions with the NFT data, the trading and minting process begins, the mechanism behind the NFT logic is " token standard" [3].

- *NFT Confirm*

Confirmation of transaction completes the minting process. And then the NFT links to a unique blockchain address as evidence.

Whenever minting of NFT is done, a new transaction is sent to invoke the smart contract. It is only after that the confirmation of transaction, the ownership details and NFT metadata are added to the newly formed block, and thus it ensures the preserve NFT history and its ownership.[3]

### 3.2 Token Standard

Token standard is the subsidiary of the smart contract standard. Token standards are included in blockchain to tell the users that how to create, issue and deploy new tokens based on their underlying blockchain.

The token standards that are used in NFT are ERC-20, ERC-721, ERC-1155. These have made a sound impact on ongoing NFT development to improve the trade of digital assets.

- *ERC-20*

This token standard introduces the concept of fungible tokens that was once satisfying the requirement of some blockchain networks. This token make these standard same as the other one, both in terms of value and type.
This means that an arbitrary token will always be equal to the other tokens present. This once invigorates the hype of Initial Coin Offering. Several public chain and various block chain based DApps gained initial funding through this [9].

- *ERC-721*

This standard introduces the non fungible token standards that are different from the fungible one introduced by ERC-20 [9]. This type of token is scarce and unique & can be differentiated from other tokens. Every NFT has a special uint256 variable called tokenId, and the pair of uint256 with the contract address is globally unique and furthermore can be used to generate special identifications [3].

- *ERC-1155*

This token standard extends the representation of both fungible and non-fungible tokens. It provides an interface that is capable of representing any number of tokens. In previous standards a token ID can contain a single type of token.For instance, ERC-20 deploys each token type in separate contracts and EC-721 deploys the group of non fungible tokens in single contract with same configuration. And then comes theERC-1155 which extends this functionality of tokenID, where every one of them can independently represent different configurable token types. The field may contain the customized information such as lock time, metadata, supply and other attributes [3].

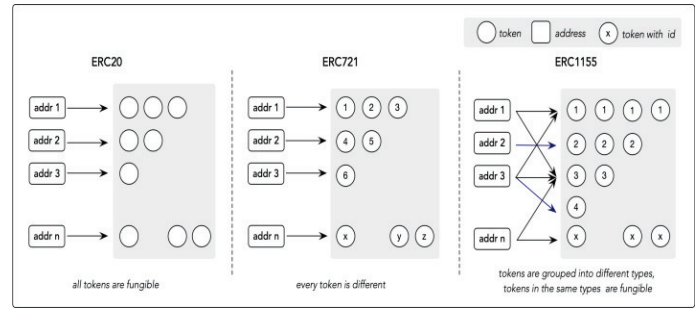

**Figure 2. Token standards of NFT [3]**

### 3.3 Key Properties

The must have properties for NFT system [3] :

- *Tradability*

NFTs and corresponding products can be randomly traded and exchanged.

- *Availability*

NFT system never goes down. It's online all the time.

- *Verifiability*

NFT has its token metadata thus its ownership can be verified publicly.

- *Usability*

NFT have information clarity as it has most up to date ownership information.

- *Tamper resistance*

When the transaction is considered affirmed and established in blockchain then, parameters like metadata of NFT and the log of records cannot be manipulated.

- *Transparent Execution*

The activities of NFT, i.e. minting, selling and purchasing are available in public domain, thus, accessible.

### 4. Security Assessment

NFT is a combination of blockchain, storage and web storage ,and because of this the security assessment of this system is a challenging task as each component of this system is a possible attacking interface and this makes the whole system

172

vulnerable to cyber attacks. Therefore, STRIDE threat and risk evaluation is adopted as it covers all facets related to security of system framework: authenticity, integrity, non-repudiability, availability and access control.

The remaining of this section comprise of potential security threats and the defense measures used against them,

- *Spoofing*

It is the ability to impersonate another entity on the structured system, which denotes vulnerable authenticity. When a user interacts to mint or sell NFT, a malicious attacker may exploit the authentication vulnerabilities. Hence the action that is prudent is to have formal verification of the smart contract and the utilization of cold wallet to forestall private key spillage.

- *Tampering*

It is the manipulation or modification in the data of NFT, which is illegal as it is not done by the user, and rather by the exploiter, and this violates integrity. As presumed it is that, the blockchain is based on public transaction ledger framework which is robust and it havehash algorithm which is preimage resistance and second preimage resistance, so the metadata and NFT ownership cannot be manipulated after the transaction is considered affirmed. But as the user are provided with the feature to store the NFT data outside the blockchain, that data can be manipulated as par with the attacker's interest. Consequently the action that is exhorted is sending both hash information just as the original data to the NFT purchaser.

- *Repudiation*

It alludes to the circumstance where the author of a statement can't question, which is identified with the security property of non-repudiability. It means that verity of established exchange of a user sending an NFT to some other user cannot be denied. This is ensured by the security of the blockchain and the unforgeability property of a signature pattern. However the hash data can be stolen or bind to the attacker's address. Thus it is believed that the multi-signature contract can somewhat address this issue since each binding must be affirmed by two or more participants. [3]

- *Information Disclosure*

The spillage of data happens when the data is presented to the unapproved users, which disregards confidentiality. The NFT framework, have state information and the instruction code stored in the smart contract are transpicuous, and any state and its amendments are open for inspection in public domain. As a matter of fact whenever a user puts NFT hash in the blockchain, the malicious aggressors of network can easily exploit the linkability of hash and transaction. And that's why it is advised to NFT developer to use the privacy-preserving smart contract.

- *Denial of Service (DOS)*

It's a type of network attack in which the assailant plans to render a server inaccessible to its preconceived user by interfering with the ordinary functionality is made to happen. This disregards the accessibility and separates the NFT service, which can be utilized by the unapproved user. New hybrid blockchain design with frail like consensus algorithm was proposed, by which this accessibility issue can be addressed. [3]

- *Elevation of Privilege*

It's a property which is identified with the approval. In this, the aggressor might acquire authorizations past those which are at first conceded. In the NFT framework the selling authorizations are overseen by the smart contract and an inadequately planned smart contract might make NFT free from such properties.

## 5. Scope

This segment lays emphasis on exploring the opportunities rising due to NFT.

- *Gaming*

NFTs have raised the potentially of booming gaming industry to an extent which is beyond measurable to its competitor sector. And NFT has given rise to new dimension in gaming, i.e. crypto gaming, there already exist some crypto games CryptoPunks, GodsUnchanged, Axie Infinity, & TradeStars. The feature of NFT that has immensely helped in changing the way games used to be , is ownership record and now because of this , the virtual collectibles and purchasable in games available as in-app purchase has seen a rise in the sales and this new technology component has been a stimuli in creation of a new ecosystem where NFT developers of such gaming items are selling their stuffs and the developers are now earning royalties each time their product is resold. [9]

- *Virtual Tickets*

The NFT has features like ownership, uniqueness, liquidity, etc. And these features can help in transforming the sectors of our physical life, for example, ticket system. Traditionally while purchasing ticket both the buyer and the seller has to trust the third party or middle man but with NFT in play, the need of third party has gone void and now the tickets can be sold to the user directly with tamperproof ownership credentials, which is not possible with the existing online ticketing system and even with the conventional ticketing there is a risk of invalid or fraudulent tickets being sold to the buyer. A NFT based ticket is one-off and scant, which implies that once the ticket is sold it can't be exchanged once more [1].

- *Transforming the Art auctioning*

Traditionally artist used to have few channels to showcase their creation.This is reflected upon the prices of their piece of art as it is not used to receive its true monetary value due to absence of attention, and with the emerging social media into the scene, still the artist were being charged the intermediary fee, which lessens the income the artists cash back [11]. But integration of NFT has transformed their art into digital format with integrated properties, and now whenever a art is purchased the ownership is transferred and due to the NFT the original creator of art can now earn royalty each the art gets resold, which was not possible earlier. Few digital artwork exchanges in the market are SuperRare, MakersPlace, VIV3, etc.[3]

## 6. Challenges
Like any nascent technology NFT also has to overcome some barriers, which are,

- *Slow confirmation of transaction*

The mechanics of NFT is that it sends the transaction to the smart contract to achieve trustworthy and transpicuous management while minting, selling. The current NFT system is closely multiplexed with the existing blockchain architecture which results in lack of performance and to solve this performance issue a new blockchain topology is needed.[3]

- *Anonymity/Privacy*

Currently the development of privacy is still being studied, the underlying Ethereum platform provides users with only pseudo-privacy and not the full privacy. But the ongoing development of zero knowledge proof provides hope that this problem will be solved in future.[1]

- *Legal seizing*

NFT is facing legal troubles like other blockchain technologies worldwide and in countries like India and China the legal situation much rigid for NFTs. The NFT trade in such countries has to overcome the governance difficulties. Legally, users in such countries can trade only on derivatives of authorized exchanges.[3]

## 7. Conclusion
NFT is a technology under development and IN this paper we have explored the NFT from the basic to the opportunities it will provide and the challenges with it, and this will make it easier for the aspiring developers and even the existing developers to keep up with the progress.